\documentclass[aps,prb,twocolumn]{revtex4-1}
\usepackage{graphicx}

\begin{document}

\title{Hydrogen inserted into the Si(100)-2$\times$1-H surface: A first-principles study}

\author{T. V. Pavlova}
 \email{pavlova@kapella.gpi.ru}
\affiliation{Prokhorov General Physics Institute of the Russian Academy of Sciences, Moscow, Russia}
\affiliation{National Research University Higher School of Economics, Moscow, Russia}

\begin{abstract}

An H atom inserted into hydrogen monolayer on the Si(100)-2$\times$1 surface has been studied using the density functional theory. Hydrogen-induced defects were considered in their neutral, negative, and positive charge states. It was found that hydrogen atom forms a dihydride unit on the surface in the most stable neutral and negative charge states. Hydrogen located in the groove between dimer rows and bonded with a second-layer Si atom is also one of the most stable negative charge states. In the positive charge state, hydrogen forms a three-center bond inside a Si dimer, Si-H-Si, similar to the bulk case. A comparison of simulated scanning tunneling microscopy (STM) images with experimental data available in the literature showed that neutral and negatively charged hydrogen-induced defects were already observed in experiments. The results reveal that the adsorption position of an H atom inserted into the Si(100)-2$\times$1-H surface is determined by the charge state of the hydrogen-induced defect.

\end{abstract}

\maketitle

\section{Introduction}

A hydrogenated Si(100) surface plays an important role in creating atomic scale devices in silicon, including single-atom transistor \cite{2012Fuechsle}, qubits on electronic spin of P atom \cite{2019He}, chains with quantum states at dangling bonds (DB) \cite{2013Schofield}, and binary logic at DB \cite{2018Huff}. To create a desired structure, hydrogen desorption lithography (HDL) in a scanning tunneling microscope (STM) is used. Today, HDL is a well-established technique with the real-atomic precision of the hydrogen removal \cite{2003Schofield, 2014Ballard, 2017Moller, 2018Achal, 2018Randall}. To control electrostatic landscape of the patterned surface, it is highly desirable to know all defects on the surface around the pattern \cite{2019Huff}. In the HDL process, most of the liberated H atoms are removed from the surface, whereas some of them interact with the surface near the lithographic pattern \cite{2014Ballard2, 2017Huff}. Indeed, a hydrogen atom can create a defect on a silicon surface because it is reactive due to an unfilled electron shell.

Hydrogen in bulk silicon is well studied due to its great technological importance, since hydrogen atoms are often embedded into silicon during semiconductor fabrication (for example, from hydride gases). Hydrogen strongly affects electrical properties of silicon, because it can passivate impurities or native defects. As a separate impurity, monoatomic hydrogen can occur in three different charge states: neutral, positive, or negative \cite{NICKEL1999}. In positive and negative charge states, hydrogen behaves as a donor or acceptor, respectively, being an amphoteric impurity \cite{2001Herring}.

Depending on the charge state, hydrogen occupies different positions in the silicon crystal. Hydrogen prefers a low electron density region in the negative charge state, and a high electron density region in the neutral and positive charge states \cite{1989VandeWalle}. In a high electron density region, hydrogen is located at the bond-center (BC) site between two Si atoms, pushing them apart by 0.4 {\AA} \cite{1991VandeWalle, 1989VandeWalle}. Since an energy cost for Si relaxation on the surface is lower than that in the bulk, the BC site inside a Si dimer on Si(100)-2$\times$1-H may be even more favorable. Besides, the Si(100)-2$\times$1-H surface has a region with a very low electron density in grooves between Si dimers, which seems suitable for a negatively charged state. Therefore, the hydrogen defects studied in bulk silicon may also exist in upper layers of the Si(100) surface. Indeed, on the similar surface, Si(100)-2$\times$1-Cl, chlorine inserted in the BC site and in the groove between dimer rows was recently observed \cite{2020PavlovaPRB}. Certainly, some defects in experimental STM images were attributed to hydrogen insertion into Si(100)-2$\times$1-H \cite{2014Ballard2, 2017Huff, 2019Huff, 2020Croshaw}. However, defect configurations and simulated STM images are still required for unambiguous interpretation of the experimental results.

The aim of this work was to study hydrogen-induced defects on the Si(100)-2$\times$1-H surface in the framework of first principles calculations. Three charge states of hydrogen in the regions with the highest and lowest electron density were considered. Surface structures of H in Si(100)-2$\times$1-H were obtained and their energy levels in the silicon band gap were calculated. The agreement with the bulk case is established only for a positive charge state. Indeed, in the positive charge state, hydrogen is favorably located inside a Si dimer at the BC site, while in neutral and negative charge states, hydrogen unites with an existed H atom to form a dihydride on top of the surface. Besides, in the groove between
dimer rows, hydrogen is also stable in a negative charge state, and its simulated STM image agrees well with STM images of the negatively charged defect described in Refs. \cite{2017Huff, 2019Huff}. In addition, realistic models are proposed for the neutral point defect and the negatively charged defect observed in STM \cite{2020Croshaw}.

\section{Calculation details}

Spin-polarized DFT calculations were carried out in the Vienna \textit{ab initio} simulation package (VASP) \cite{1993Kresse, 1996Kresse} by employing the projected augmented wave method \cite{1994Blochl}. The exchange and correlation effects in the electron gas were treated within the generalized gradient approximation of Perdew, Burke, and Ernzerhof \cite{1996Perdew}. The van der Waals correction was included by using the DFT-D2 method developed by Grimme \cite{2006Grimme}. Plane wave cutoff energy was set at 400 eV. The Si(100)-2$\times$1 surface was modeled by an eight layer slab with a 5$\times$6 supercell. Hydrogen atoms were placed on the upper surface to form a Si(100)-2$\times$1-H structure, whereas DBs on the bottom surface were saturated by dihydride structure. The bottom three Si layers were fixed at their bulk equilibrium positions, while the other Si and upper H atoms were allowed to relax. To avoid interaction between the neighboring layers, the slabs were separated by a vacuum space of 21\,{\AA}. The reciprocal cell was integrated using a 4$\times$4$\times$1 k-point grid. The electronic density of states (DOS) was evaluated using a 16$\times$16$\times$1 k-point grid. According to Ref. \cite{2017Scherpelz}, energies of DB levels within the band gap depend significantly on the slab thickness. Therefore, the parameters used in this work allow only a qualitative assessment of the level positions, since DOS calculations converge slow with respect to the slab thickness \cite{2017Scherpelz}. To simulate positive and negative charge states, the electron was removed or added to the supercell, respectively, that is a well-known method for modeling charge states on silicon surfaces \cite{2013Schofield, 2012Studer, 2006Radny}. STM images were generated in the HiveSTM program \cite{2008Vanpoucke} using the Tersoff-Hamann approximation \cite{1985Tersoff}. The adsorption energies of hydrogen were calculated as the difference between the total energy of the Si(100)-2$\times$1-H surface with the H atom and the total energies of the surface and the H atom in the gaseous phase.

\section{Results and discussion}

\begin{figure}[h]
 \includegraphics[width=\linewidth]{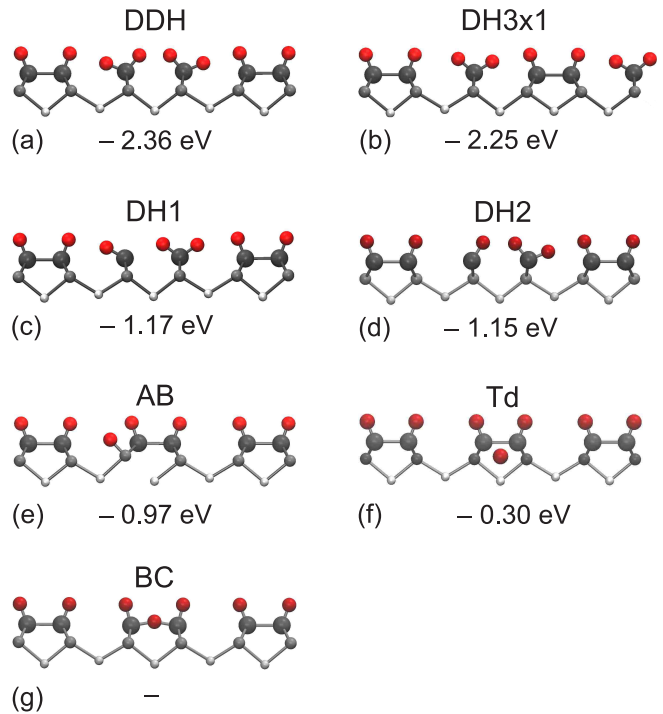}
\caption{\label{fig1} Side views of the Si(100)-2$\times$1-H surface with two (a),(b) and one (c)-(g) H atom added to the supercell in different sites. (a) Double dihydride on a Si dimer (DDH). (b) The (3$\times$1) local structure inside the Si(100)-2$\times$1-H surface (DH3x1). (c) Dihydride on a Si dimer (DH1). (d) Same as (c), but the H atom of the SiH unit is turned (DH2). (e) Hydrogen at the antibonding site on the Si-Si bond extension (AB). (f) Hydrogen at the tetrahedral interstitial site (Td). (g) Hydrogen at the bond-center site (BC). Hydrogen atoms are marked in red, silicon atoms in gray. Adsorption energies are given below the structures.}
\end{figure}

A Si(100) monohydride phase is reconstructed (2$\times$1) with a Si dimer structure, in which a DB at a Si atom is terminated by an H atom. Along with the monohydride phase, the dihydride phase and the (3$\times$1) reconstruction are usually presented on the surface \cite{2008Bellec}. In the local dihydride structure, two H atoms inserted into Si(100)-2$\times$1-H form two SiH$_2$ units on one dimer. In this structure, two adjacent SiH$_2$ are canted in opposite directions (Fig.~\ref{fig1}a). The local (3$\times$1) structure also contains two additional H atoms forming two SiH$_2$ units, whereas two Si-Si dimer bonds are broken and a new one is formed in the middle (Fig.~\ref{fig1}b).

At singe H atom insertion into the Si(100)-2$\times$1-H surface, one SiH$_2$ unit can form (Fig.~\ref{fig1}c,d). Figure~\ref{fig1}e-g shows another adsorption positions of hydrogen, similar to those in bulk silicon \cite{1991VandeWalle}, but near the surface. In the antibonding (AB) configuration, H is located on the extension of a Si-Si bond in a groove between dimer rows, where the crystal charge density is low (Fig.~\ref{fig1}e). Another low density region near the surface is tetrahedral (Td) interstitial site between two dimers (Fig.~\ref{fig1}g). In the high density region in the upper surface layer, hydrogen is placed midway between two Si atoms of the dimer, in the BC site (Fig.~\ref{fig1}g).

Adsorption energies of the configurations are given in Fig.~\ref{fig1}, except for BC, which is not stable in a neutral supercell. Configurations DDH and DH3x1 with two inserted H atoms are more stable than the configurations with one inserted H atom by more than 1 eV. Indeed, dihydride and the (3$\times$1) reconstruction are the most often observed defects on the Si(100)-2$\times$1-H surface. However, the insertion of a single hydrogen atom into the surface also gives a gain in energy since the adsorption energy is negative.

\begin{figure}[h]
 \includegraphics[width=\linewidth]{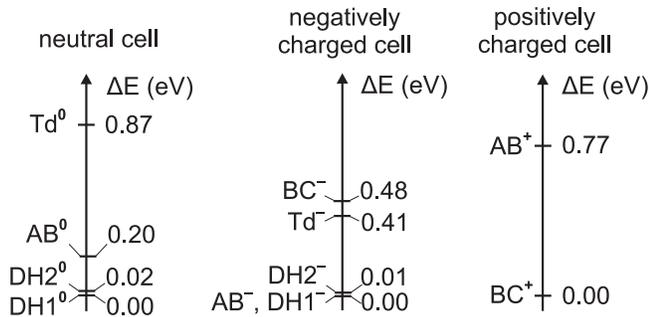}
\caption{\label{fig2} Relative total energies for configurations DH1, DH2, AB, Td, and BC in neutral, negative, and positive charge states. The zero of energy corresponds to the energy of the most stable configuration.}
\end{figure}

Relative energies for all considered configurations with a single H atom in the neutral, negative, and positive charge states are given in Fig.~\ref{fig2}. The use of notation, for example, BC$^+$ (BC$^-$), emphasizes that an electron was removed (added) from (to) the defect complex but not only from (to) the H atom. Not all the configurations are stable in all three charge states. Specifically, positively charged DH1 and DH2 configurations relax into the BC$^+$ configuration. Positively charged Td turns into the BC site with the H atom in between second- and third-layer Si atoms. Neutral BC configuration converts to DH1$^0$, although BC is the most stable site in the neutral charge state in a bulk \cite{1989VandeWalle}.

The stability of different charge states determines the preferable position of hydrogen on Si(100)-2$\times$1-H.
In a neutral state, hydrogen prefers to form a dihydride (DH1$^0$ and DH2$^0$). A negative charge state is most stable in the low electron density region: above the surface (DH1$^-$ and DH2$^-$) and in the groove between the dimer rows (AB$^-$), although the lowest electron density region in a bulk is in the Td site. Thus, only BC$^+$  is the most stable positively charged configuration both on the surface and in a bulk.

\begin{figure*}[t]
\includegraphics[width=\linewidth]{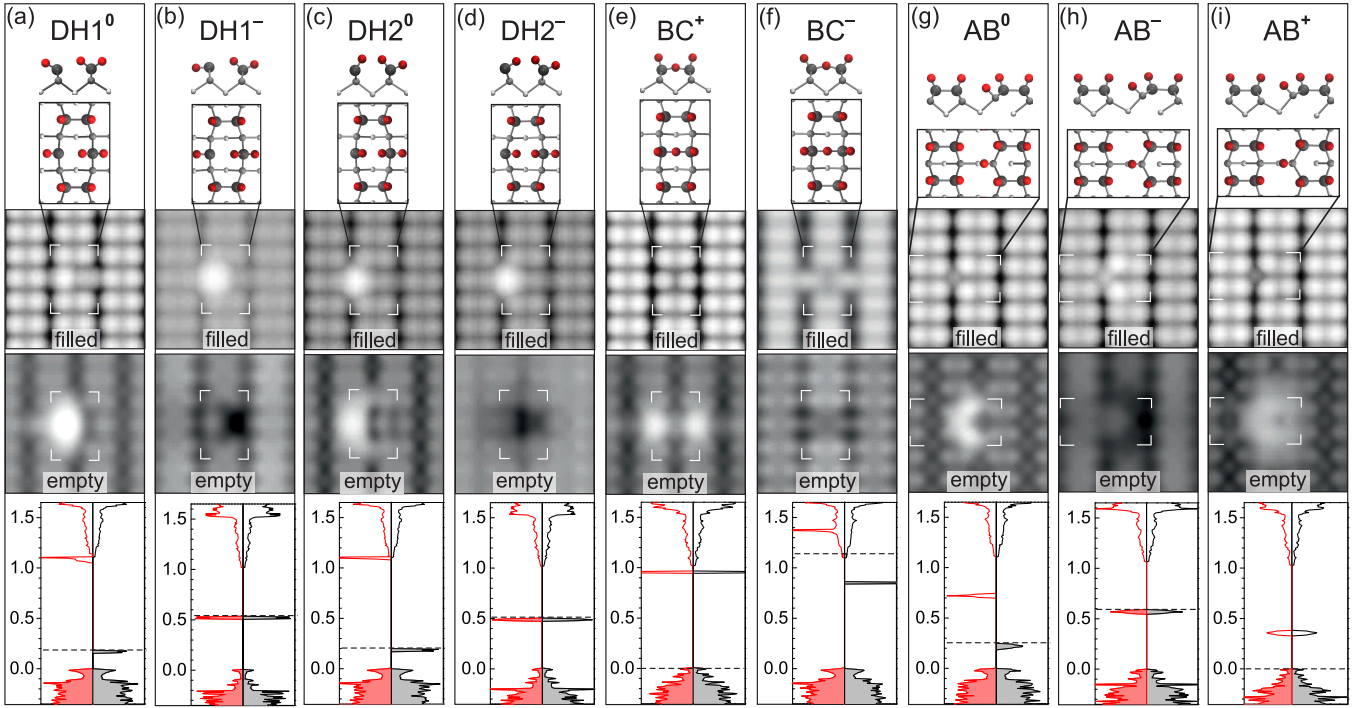}
\caption{\label{fig3} Hydrogen configurations on Si(100)-2$\times$1-H in different charged states: (a) DH1$^0$, (b) DH1$^-$, (c) DH2$^0$, (d) DH2$^-$, (e) BC$^+$, (f) BC$^-$, (g) AB$^0$, (h) AB$^-$, (i) AB$^+$. Side and top views of models, simulated filled- ($-$1.8\,V) and empty- ($+$1.3\,V) state STM images (empty-state STM image of DH2$^-$ was simulated at $+$1.15\,V), and total DOS in the vicinity of the band gap are shown from top to down. The energies in the DOS plots are given in eV relative to the VBM. Spin-up and spin-down states are shown in red and black, filled states are filled with color. The Fermi level is shown with the dashed line. Hydrogen atoms are marked in red, silicon atoms in gray.}
\end{figure*}

Figure~\ref{fig3} shows stable configurations with one inserted H atom (excluding Td, which is disadvantageous in any charge state): models, STM images, and DOS. In DH1$^0$, the Si-Si bond inside the dimer is broken, so the Si atom of the SiH unit has a DB (Fig.~\ref{fig3}a). In DOS, the DB has two states in the band gap: occupied and unoccupied \footnote{ Note that the DB level below the conduction-band minimum converges well to the slab thickness, while the DB level at the bottom of the band gap may fall slightly below the VBM \cite{2017Scherpelz}.}, similar to the case of a DB on Si(100)-2$\times$1-H with a H vacancy \cite{2017Scherpelz}. The unoccupied state in the band gap is visualized as a bright lobe in the empty-state STM image, as in the case of the H vacancy \cite{2016Kawai}. In the work of Ballard et al. \cite{2014Ballard2}, configuration DH1$^0$ was observed experimentally and also identified as a DB on an unsaturated dihydride dimer.

An additional electron of DH1$^-$ is localized on the DB at the Si atom of the SiH unit (Fig.~\ref{fig3}b). The structural model of DH1$^-$ differs from that of DH1$^0$ due to Si orbitals hybridization in sp$^3$. Indeed, according to the DOS plot, both DB states in the band gap are occupied. Note that in the empty-state STM image (Fig.~\ref{fig3}b), the SiH$_2$ unit is visualized as a depression.

\begin{figure}[h]
 \includegraphics[width=\linewidth]{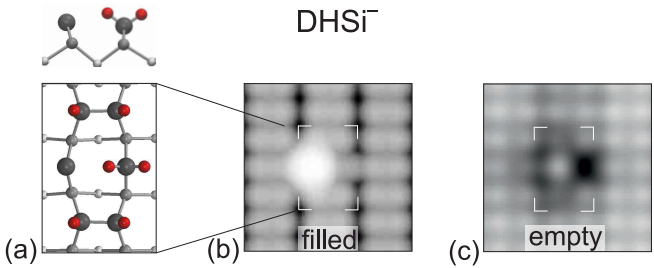}
\caption{\label{fig4} Negatively charged DHSi$^-$ configuration on the Si(100)-2$\times$1-H surface. (a) Side and top views of the model; (b) simulated filled- ($-1.8$\,V) and (c) empty-state ($+1.8$\,V) STM images.}
\end{figure}

According to the published STM data, DH2$^0$ (Fig.~\ref{fig3}c) was not observed, while DH2$^-$ (Fig.~\ref{fig3}d) is suitable for identifying the so-called ``negatively charged defect'' \cite{2020Croshaw}. The empty-state STM images of DH2$^-$ was simulated at $+$1.15\,V to compare it with experimental STM images \cite{2020Croshaw} in which the electron density is combined between an H atoms in different dimer rows. This effect is reproduced at a lower voltage when less unoccupied states are taken into account. Interestingly, the ``negatively charged defect'' converts into the so-called ``neutral point defect'' at the hydrogen removal \cite{2020Croshaw}. A model of the DHSi$^-$ configuration formed by the hydrogen removal from the SiH unit of DH2$^-$ is shown in Fig.~\ref{fig4}a. The simulated STM images of DHSi$^-$ are similar to the experimental ones of the ``neutral point defect'' \cite{2020Croshaw}: a bright lobe and a bright dot in the filled- (Fig.~\ref{fig4}b) and empty-state (Fig.~\ref{fig4}c) STM images, respectively. By contrast, in the neutral DHSi configuration, the unoccupied DB states in the band gap are visualized as a big bright lobe in the empty-state STM image, which is not consistent with the experiment \cite{2020Croshaw}. Therefore, the DHSi$^-$ configuration is negatively charged. Thus, we believe that the negatively charged configurations DH2$^-$  and DHSi$^-$ were observed in Ref.~\cite{2020Croshaw} as the ``negatively charged defect'' and ``neutral point defect'', respectively.

In configuration BC$^+$ (Fig.~\ref{fig3}e), the Si-H-Si bond is almost linear, and the Si–H distance is equal to 1.66 {\AA}. To accommodate the hydrogen atom, silicon atoms move outward to a distance of 3.27 {\AA}, which indicates significant relaxation compared to the unperturbed dimer (2.41\,{\AA}). In the band gap, configuration BC$^+$ has a shallow donor level, as in the bulk case \cite{NICKEL1999}. Despite the neutral configuration BC$^0$ is not stable, negatively charged BC$^-$ has a local minimum with the Si-H bond length of 1.75 {\AA} (Fig.~\ref{fig3}f).

Although BC$^+$  is the most stable positively charged configuration (Fig.~\ref{fig2}), no observation of this state has been reported. One of the possible ways to create BC$^+$ is to remove an electron from the occupied state of DH1$^0$ (DH2$^0$), which was observed earlier \cite{2014Ballard2}. Indeed, if the electron is removed, configuration DH1$^0$ (DH2$^0$) converts into BC$^+$. Changing of a DB charge state was already demonstrated using STM \cite{2013Bellec} and noncontact atomic force microscope (nc-AFM) \cite{2018Rashidi}. Most easily an electron may be removed in a p-doped sample at an upward tip induced band banding (empty-state STM-imaging). At a negative voltage bias, BC$^+$ may convert back to DH1$^0$ (DH2$^0$) or DH1$^-$ (DH2$^-$), if one or two electrons are added, respectively. Note that a similar three-center configuration with chlorine, Si-Cl-Si, was recently observed on the Si(100)-2$\times$1-Cl surface \cite{2020PavlovaPRB}, and this configuration converts into configuration AB$^-$ at the changing of the voltage polarity from positive to negative.

Configuration  AB$^0$ (Fig.~\ref{fig3}g) has a DB at a third-layer Si atom, and therefore the band gap contains two DB states, similar to DB states on the Si(100)-2$\times$1-H surface with an H vacancy \cite{2017Scherpelz}. In AB$^-$ (Fig.~\ref{fig3}h) and AB$^+$ (Fig.~\ref{fig3}i), both states are occupied or unoccupied, respectively. In AB$^-$, the angles between the third-layer Si atom and three neighboring atoms are 105$^\circ$, which indicates sp$^3$-like hybridization of the orbitals of the third-layer Si atom. In contrast, the angles in AB$^+$ are 118$^\circ$, which pointed out to sp$^2$-like hybridization.

The empty-state STM image of AB$^-$ appears similar to the STM images of the so-called ``negative hydrogen'' on the Si(100)-2$\times$1-H surface observed in Refs.~\cite{2017Huff, 2019Huff}. On the simulated and experimental empty-state STM images, a dark halo is visible around a brighter region. It is well known that the halo arises around a negatively charged DB due to the charge induced band banding \cite{2013Schofield, 2015Labidi}. Therefore, the halo around AB$^-$ should be attributed to the electron localization on the DB at the third-layer Si atom. In nc-AFM images \cite{2017Huff, 2019Huff}, an H atom is not visible, and the nearest two H atoms are pulled apart. The model AB$^-$ (Fig.~\ref{fig3}h) looks exactly like the nc-AFM images \cite{2017Huff, 2019Huff}, with H located 1 {\AA} below the surface and therefore invisible. In the previous works \cite{2017Huff, 2019Huff}, the H atom in a groove between two adjacent dimer pairs was considered as the physisorbed H atom with a negative charge. However, two H atoms pulled apart are the direct consequence of the Si-H bond formation between a second-layer Si atom and hydrogen. Formation of the Si-H bond and rather high adsorption energy about $-1$ eV for AB$^0$ indicate the chemisorption state. Thus, a lone hydrogen atom observed in the experiments \cite{2017Huff, 2019Huff}, is indeed located in the groove between two adjacent dimer pairs, but it is chemisorbed rather than physisorbed. Additionally, H is not actually the ``negative hydrogen'', because the electron is added mostly to the DB at the third-layer Si atom but not solely to the H atom.

\begin{figure}[h]
 \includegraphics[width=\linewidth]{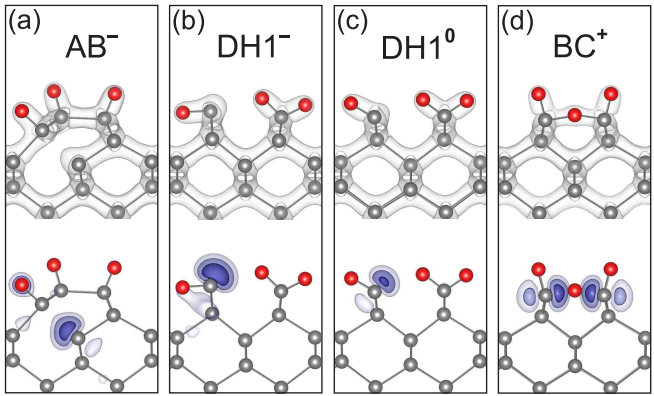}
\caption{\label{fig5} Cross-sections of the charge distribution in the most stable configurations: (a) AB$^-$, (b) DH1$^-$, (c) DH1$^0$, and (d) BC$^+$. Top panel: the electron charge density with isosurface level of 0.05 e/{\AA}$^3$; bottom panel: ILDOS with isosurface levels of 0.005, 0.010, and 0.020 e/{\AA}$^3$ (from light to dark blue) for the band gap states. In (a)-(c), the ILDOS is calculated for the occupied state, and in (d) the ILDOS is calculated for the unoccupied state.}
\end{figure}

To demonstrate the charge localization on the DB in AB$^-$, the integrated local density of states (ILDOS) of the occupied state in the band gap was calculated (Fig.~\ref{fig5}a). Indeed, the electron is localized not so much on the H atom, as on the DB at the third-layer Si atom. In addition, the bond breaking between the second- and third-layer Si atoms is confirmed in the electron charge density distribution (Fig.~\ref{fig5}a). In both DH1$^-$ (Fig.~\ref{fig5}b) and DH1$^0$ (Fig.~\ref{fig5}c) configurations, the occupied level in the band gap corresponds to the localized DB state at the Si atom of the SiH fragment. In BC$^+$ (Fig.~\ref{fig5}d), the H atom is located in a high electron density region, at a node of an antibonding combination of Si orbitals. Thus, in BC$^+$, an electron is removed from this antibonding combination but not from the H atom itself, as it was shown for the positively charged Si-H-Si complex using a simple molecular-orbital picture \cite{1989VandeWalle}. Therefore, the charge in AB$^-$, DH1$^-$, and BC$^+$ is actually not localized on the H atom but belongs to the whole configuration.

\section{Conclusions}

The first-principles calculations demonstrate that inserted hydrogen is stable on the Si(100)-2$\times$1-H surface in three charge states: neutral, positive, and negative. The type of the charge state defines the most favorable adsorption site of an H atom. In a positively charged supercell, the H atom locates in the middle of a Si dimer (BC$^+$). In a neutral supercell, the formation of a SiH$_2$ unit (DH1$^0$) on the surface is more preferable. Hydrogen in the SiH$_2$ unit (DH1$^-$) and in the groove between dimer rows (AB$^-$) are the most stable configurations in a negatively charged supercell.

The simulated STM images of inserted hydrogen provide new insight into the observed earlier defects on the Si(100)-2$\times$1-H surface \cite{2017Huff, 2019Huff, 2020Croshaw}. The ``negatively charged defect''  \cite{2020Croshaw} is now interpreted as DH2$^-$ configuration. At hydrogen removal, ``negatively charged defect'' converts into ``neutral point defect'' \cite{2020Croshaw}, which is assigned in the present work to the negatively charged DHSi$^-$ configuration. ``Negative hydrogen'' observed in the STM experiments \cite{2017Huff, 2019Huff} is now interpreted as a chemisorbed hydrogen in configuration AB$^-$ with a negative charge located mainly on the DB at the third-layer Si atom.

To insert hydrogen into the Si(100)-2$\times$1-H surface, an H-functionalized STM tip can be used. The technique of hydrogen deposition on the H-vacancy site on Si(100)-2$\times$1-H from the H-functionalized STM tip is well established \cite{2017Niko, 2017Huff, 2018Achal}. The same technique may lead to hydrogen insertion into the defectless Si(100)-2$\times$1-H surface, as it was done for the Cl insertion into the Si(100)-2$\times$1-Cl surface from the Cl-functionalized STM tip \cite{2020PavlovaPRB, 2020Pavlova}. It would allow the controllable creation of the defects considered in this work. For example, the DB at the third-layer Si atom can be of interest for quantum engineering with DB since it is better protected from the adsorption of molecules compared to a DB on top of the surface. It should, however, be pointed out that hydrogen in configuration AB$^-$ is more mobile than the hydrogen vacancy on the Si(100)-2$\times$1-H surface.

Today, automatic identification of surface defects on the Si(100)-2$\times$1-H surface using machine learning is actively developed  \cite{2020Rashidi, 2020Ziatdinov, 2020Usman}. To train a machine learning algorithm, a library of STM images of defects is usually required. Since many defects studied in this work were already observed \cite{2014Ballard2, 2017Huff, 2019Huff, 2020Croshaw}, they should be included in the corresponding library. In addition, further work is needed to consider hydrogen-induced defects in a deeper subsurface region.

\begin{acknowledgments}
This study was supported by the Russian Science Foundation under the grant No. 16-12-00050. All calculations were performed at the Joint Supercomputer Center of RAS.
\end{acknowledgments}

\bibliographystyle{apsrev4-1}
\bibliography{Pavlova_H_arxiv}

\end{document}